\documentclass[10pt,letterpaper,journal]{IEEEtran}
\pdfoutput=1
\usepackage[T1]{fontenc}
\usepackage[utf8]{inputenc}
\usepackage{graphicx}
\usepackage{cite}
\usepackage[pdftex,
            pdfauthor={Axel Huebl},
            pdftitle={Visualizing the Radiation of the Kelvin-Helmholtz
                      Instability},
            pdfsubject={IEEE Special Issue on Images 2014},
            pdfkeywords={Plasma stability, Simulation, Parallel machines,
Particle tracking, Electromagnetic radiation, Visualization}]{hyperref}
\usepackage{authblk}

\title{Visualizing the Radiation of the Kelvin-Helmholtz Instability}

\author{Axel~Huebl, David~Pugmire, Felix~Schmitt, Richard~Pausch and Michael~Bussmann%
\thanks{Manuscript received November 8, 2013; revised March 3, 2014}%
\thanks{This research used resources of the Oak Ridge Leadership Computing Facility at the Oak Ridge National Laboratory,
which is supported by the Office of Science of the U.S. Department of Energy under Contract No. DE-AC05-00OR22725.}%
\thanks{A.~Huebl (a.huebl@hzdr.de), R.~Pausch and M.~Bussmann are with the Helmholtz-Zentrum Dresden - Rossendorf}%
\thanks{D.~Pugmire is with the Oak Ridge National Laboratory}%
\thanks{F.~Schmitt is with Dresden University of Technology}%
\thanks{This work is submitted for possible publication to the 7th Triennial Special Issue of the IEEE Transactions on Plasma Science, Images in Plasma Science.
The IEEE copyright notices apply.}}

\date{\today}

\begin{document}

\maketitle

%%%%%%%%%%%%%%%%%%%%%%%%%%%%%%%%%%%%%%%%%%%%%%%%%%%%%%%%%%%%%%%%%%%%%%%%%%%%%%%

\begin{abstract}
Emerging new technologies in plasma simulations allow tracking billions of particles while computing their radiative spectra.
We present a visualization of the relativistic Kelvin-Helmholtz Instability from a simulation performed with the fully relativistic particle-in-cell code PIConGPU powered by 18,000 GPUs on the USA’s fastest supercomputer Titan~\cite{TOP500}.
\end{abstract}

%%%%%%%%%%%%%%%%%%%%%%%%%%%%%%%%%%%%%%%%%%%%%%%%%%%%%%%%%%%%%%%%%%%%%%%%%%%%%%%

How do we see what happens in jets emanating from active galactic nuclei or gamma-ray bursts~\cite{Colgate01}, if we cannot resolve the plasma structures with our telescopes?
One phenomenon of interest in hot plasma jets spewing out into space is the Kelvin-Helmholtz Instability (KHI).
It occurs at the interface of two plasma streams that flow at different speeds, causing a turbulent, chaotic mix of the two streams.
The KHI, long time investigated using fluid and hybrid simulations~\cite{Angelo65}~\cite{Thomas93}, could only very recently be studied using particle-in-cell (PIC) simulations~\cite{Alves12}.

\begin{figure}[ht]
    \centering
    \includegraphics[width=0.95\textwidth]{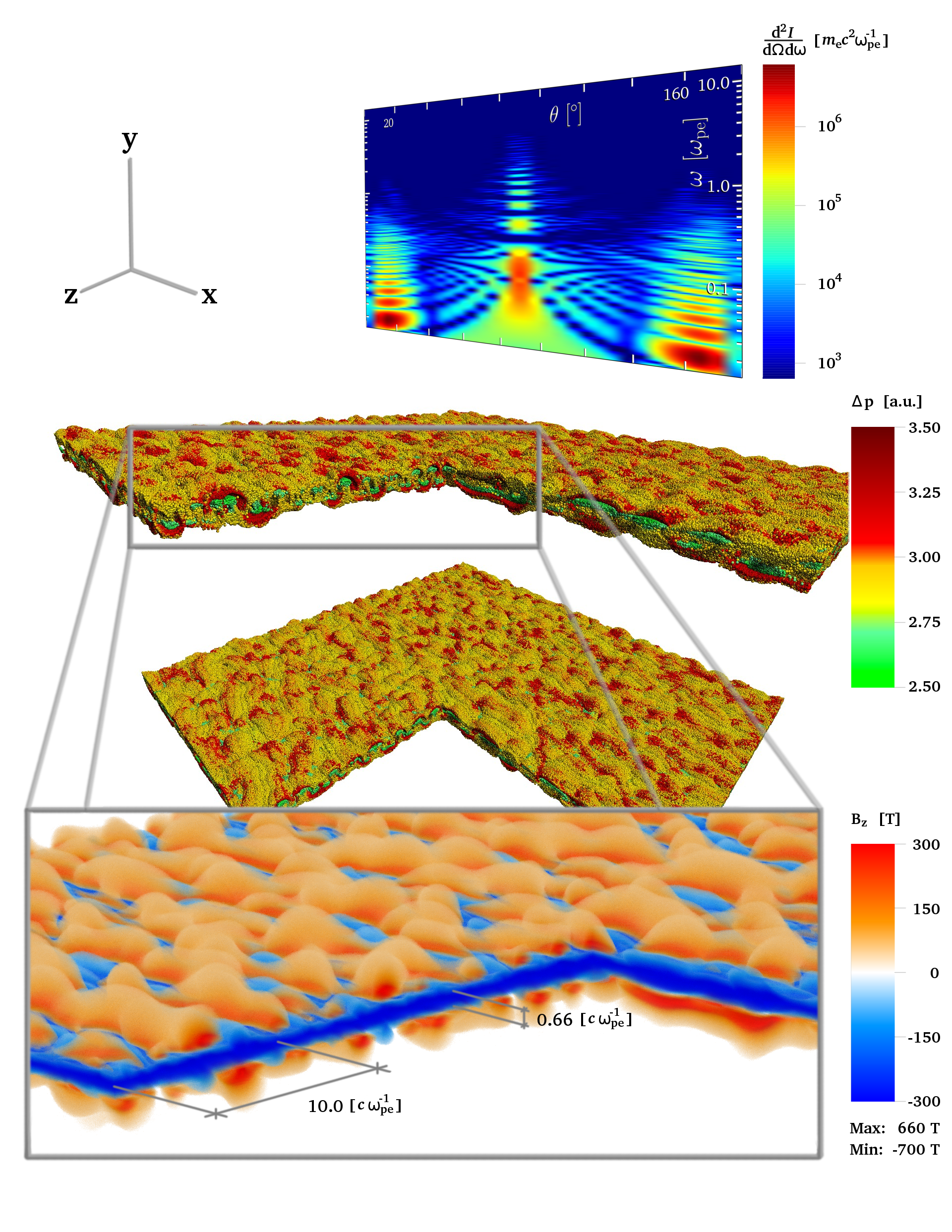}
    \caption{(Center) Force-picture of the electrons at time $t_{900} = 27 / \omega_{pe^-}$. In order to reduce the noise caused by thermal pertubations we filtered by momentum, selecting particles with $p_y\vee p_z \ge 0.05 \cdot p_x$.
        (Bottom) The magnetic field $B_z$ generated by the currents at the shear surface.
        (Top) Section of the spectral sky map scanning in the x-y plane. The observation direction is given by the angle $\theta$ being spanned from the negative x towards the y axis.
        (Computational domain: $480\times46\times46$ $c/\omega_{pe^-}$, $8000\times768\times768$ cells, 16 macro particles (8 for each species) per cell, e$^-$ to p$^+$ mass ratio $1:1836$, reference $n_{e^-}=10^{25}$ m$^{-3}$ for presented units, simulated $75.5\cdot10^{9}$ particles in total)}
    \label{fig1}
\end{figure}

We present a visualization of a fully relativistic PIC simulation of the KHI performed on the Titan supercomputer~\cite{Bussmann13}.
Achieving a peak performance of 7.2\,PFlop/s, this simulation raises the standard of today's kinetic plasma simulations.
Unleashing the computational power of 18,000 GPUs we simulated the KHI in a volume 46 times larger with 4.2 times more spatial resolution than any existing simulation before~\cite{Alves12} (for simulation parameters see caption of Fig. \ref{fig1}).
Moreover, we computed synthetic radiation spectra~\cite{Pausch13} from the motion of 19 billion electrons.
From this we produced a spectral sky map, scanning 481 directions and 512 frequencies for each direction, ranging from 0.014 to 14 times the plasma frequency.

The presented setting was given by two preionized, unmagnetized hydrogen plasma slabs drifting in $\pm x$ direction inside a simulation box with periodic boundary conditions, with an initial relativistic gamma factor of 3 each.
The KHI builds up independently on both shear surfaces triggered ab initio by thermal noise.
While electrons get scattered in the counter propagating stream a current imbalance is induced.
Self-consistently generated magnetic fields (see inset $B_z$) in turn drive the particle dynamics creating a feedback loop~\cite{Alves10}.

The image presented here shows the electron dynamics, magnetic field structures and the radiation spectra of electrons that can be observed in the far-field.
It shows strong local changes in the electron momenta and local magnetic fields.
An analysis of the corresponding radiation spectra in the far-field is ongoing.

The simulation was performed with PIConGPU, an open source many GPGPU code written in C++/CUDA and published under GPLv3+~\cite{picongpu}.
The authors would like to thank the Oak Ridge National Laboratory and the Center for Information Services and High Performance Computing, Dresden University of Technology for the provided support.
We thank all PIConGPU developers for their efforts to make this possible.

%%%%%%%%%%%%%%%%%%%%%%%%%%%%%%%%%%%%%%%%%%%%%%%%%%%%%%%%%%%%%%%%%%%%%%%%%%%%%%%

\bibliographystyle{IEEEtran}
\bibliography{IEEEabrv,refs}

%%%%%%%%%%%%%%%%%%%%%%%%%%%%%%%%%%%%%%%%%%%%%%%%%%%%%%%%%%%%%%%%%%%%%%%%%%%%%%%

\onecolumn

%%%%%%%%%%%%%%%%%%%%%%%%%%%%%%%%%%%%%%%%%%%%%%%%%%%%%%%%%%%%%%%%%%%%%%%%%%%%%%%

\end{document}